\documentclass[conference]{IEEEtran}

\IEEEoverridecommandlockouts                              

\usepackage{cite}
\def\BibTeX{{\rm B\kern-.05em{\sc i\kern-.025em b}\kern-.08em
    T\kern-.1667em\lower.7ex\hbox{E}\kern-.125emX}}
    
\usepackage{amsmath}
\usepackage{bm}
\usepackage{amsthm}
\usepackage{amssymb}
\usepackage{amsfonts}
\usepackage{float}
\usepackage{soul}
\usepackage{algorithmic}
\usepackage{tensor}
\usepackage{graphicx}
\usepackage{physics}
\usepackage{textcomp}
\usepackage{wrapfig}
\usepackage{caption}
\usepackage{hyperref}
\usepackage{subcaption}
\usepackage{multirow}
\usepackage{comment}
\usepackage{steinmetz}
\usepackage{mathtools}
\usepackage{cuted}

\usepackage[table]{xcolor}
\theoremstyle{definition}
\newtheorem{remark}{Remark}

\title{\LARGE \bf
The Role of Time Delay in Sim2real Transfer of Reinforcement Learning for Cyber-Physical Systems}

\author{
        Mohamad~Chehadeh\href{https://orcid.org/0000-0002-9430-3349}{\includegraphics[scale=0.75]{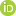}},~\IEEEmembership{Member,~IEEE,}\\
        Igor~Boiko\href{https://orcid.org/0000-0003-4978-614X}{\includegraphics[scale=0.75]{figures/orcid.png}},~\IEEEmembership{Senior~Member,~IEEE}
        and~Yahya~Zweiri\href{https://orcid.org/0000-0003-4331-7254}{\includegraphics[scale=0.75]{figures/orcid.png}},~\IEEEmembership{Member,~IEEE}
\thanks{This work was supported by Khalifa University Grant CIRA-2020-082.

M. Chehadeh, I. Boiko, and Y. Zweiri are with the Center for Autonomous Robotic Systems, Khalifa University, Abu Dhabi, United Arab Emirates. Also, I. Boiko is with the Department of Electrical Engineering and Computer Science, and Y. Zweiri with the Department of Aerospace Engineering, both at Khalifa University, Abu Dhabi, United Arab Emirates.}
}

\begin{document}

\maketitle
\thispagestyle{empty}
\pagestyle{empty}

\begin{abstract}
This paper analyzes the simulation to reality gap in reinforcement learning (RL) cyber-physical systems with fractional delays (i.e. delays that are non-integer multiple of the sampling period). The consideration of fractional delay has important implications on the nature of the cyber-physical system considered. Systems with delays are non-Markovian, and the system state vector needs to be extended to make the system Markovian. We show that this is not possible when the delay is in the output, and the problem would always be non-Markovian. Based on this analysis, a sampling scheme is proposed that results in efficient RL training and agents that perform well in realistic multirotor unmanned aerial vehicle simulations. We demonstrate that the resultant agents do not produce excessive oscillations, which is not the case with RL agents that do not consider time delay in the model.
\end{abstract}

\section{Introduction}
The use of data-based approaches for control has been attracting increased attention recently from the robotics community. Data-based approaches promise to offer advantages over the classical model-based control approaches in some aspects; mainly in the handling of complex or impossible to model control problems. Reinforcement Learning (RL) is the most prominent of these data-based approaches and its usage witnessed a surge in the robotics community following the notable success in other domains like computer games \cite{mnih2015human}, in the hope that the same benefits would be claimed with real robots. Unfortunately the application of RL in robotics was not as successful, mainly due to the discrepancy between the simulation models used for RL training, and the real robots.

\subsection{Related Work}
Successful transfer from simulation to reality (S2R) largely depends on the validity of the simulator in use \cite{ibarz2021train}. A few approaches has been proposed to close the S2R gap like domain randomization (DR) in simulation \cite{tobin2017domain}, and domain adaptation (DA) where simulation data is augmented with real-world data to improve model fidelity used in the training of RL \cite{chebotar2019closing}. The methods that fall under DR can be categorized into two categories: dynamics randomization \cite{Peng2018}, and random perturbations of states and actions. It was found in \cite{ibarz2021train} that the mere addition of perturbations to the observed states or actions in simulation is not enough to improve S2R transferability. Dynamics randomization increases the robustness of the transferred policies, yet it significantly increases training time and complexity, and in some cases physically unrealizable dynamics were found to improve transferability like including a negative mass in the randomization range \cite{margolis2022rapid}. It was observed that DA improves transferability of policies as real world data is used to either improve a generative simulation model or directly for agent training. In some sense DA can be seen as a system identification tool that improves the simulation model \cite{chebotar2019closing}.

Another reason for the S2R gap is the omission of actuator dynamics and delays in the system. These were observed mainly in the design of RL controllers for agile locomotion of quadruped robots \cite{ibarz2021train,tan2018sim,haarnoja2018learning}, and multirotor UAVs \cite{lambert2019low,kaufmann2022benchmark}. The omission of actuator and delay dynamics would make the problem partially observable (i.e. non-Markovian) which would make, in these circumstances, every other RL technique like DR and DA hard to apply and to reason about. 

The partially observable Markov decision process (POMDP) can be made close to an MDP by considering previous states and actions \cite{haarnoja2018learning}. In exact theoretical sense, it is possible to transform any POMDP to a MDP by expanding the states vector with actions and observations in the cases of deterministic \cite{katsikopoulos2003markov,chen2021delay,schuitema2010control} and random \cite{bouteiller2020reinforcement} delays. The limitation of these approaches is that they assume a delay that is an integer step of the sampling period, they assume the knowledge of such delay or its distribution, or they assume perfect model that can be used for future states prediction. Such assumptions yield approximate simulation models with required priors that are not accessable in practice.

The authors argued in \cite{ibarz2021train} that the causes of the S2R gap are the incorrect physical parameters, the unmodeled dynamics, and the stochastic real environment. In \cite{ibarz2021train} it is pointed out that there is no general consensus about which of these causes play the most important role in the S2R gap. Our recent theoretical analysis in \cite{humais2022analysis} showed that considering delay in the models used for the tuning of UAV controllers is essential for the controller to transfer directly to experimentation. Thus we argue in this work that unmodeled dynamics are the main contributors to the S2R gap.

\subsection{Contribution}
The contribution of this work is the consideration of fractional delays (i.e. delays that are a fraction of the sampling period) in the dynamic model used for RL agent training. The dynamic model corresponds to the altitude and attitude dynamics of multirotor UAVs, and accounts for actuator dynamics and system delays (i.e. delays in the input or output) \cite{chehadeh2019design,humais2022analysis}. We show that the POMDP can be transferred to MDP in the case of input delay, but requires a model based estimator if the delay is in the output. We also show that considering the detailed model in RL design yields agents that have different qualities compared to delay-free models, most notably the fact that RL agents trained on delay-free models result in oscillatory behaviour when applied to the delayed model which explains the behaviour reported in \cite{tan2018sim}. The testing simulation environment we use is believed to accurately represent real dynamics since it was used to experimentation predict stability limits in \cite{humais2022analysis}, and it also predicts the system sensitivity to delay changes \cite{Ayyad2020}.

\section{UAV Altitude Model}
We chose to perform our testing on a UAV altitude model that accounts for actuator and delay dynamics. The delay free model that includes first order actuator dynamics and aerodynamic drag is given by \cite{chehadeh2019design} (note that we assume gravity compensation):
\begin{equation}\label{eq_state_space}
\begin{split}
\mathbf{\dot{x}}_z &= A_z \mathbf{x}_z +B_z u_z\\
\mathbf{y}_z       &= C_z \mathbf{x}_z + D_z u_z 
\end{split}
\end{equation}
where the system matrices \(A, B, C, D\) are given by:

\begin{gather*}
A =
  \begin{bmatrix}
   0 & 1 & 0 \\
   0 & 0 & 1 \\
   0 & \frac{-1}{T_p T_q} & -\frac{T_p+T_q}{T_p T_q}
   \end{bmatrix}, \\
B=
   \begin{bmatrix}
       0 \\ 0 \\ \frac{K_z}{T_p T_q}
   \end{bmatrix},
   \\
   C=
   I_{3}, \\
   D=0
\end{gather*}
where \(T_p\) is the time constant associated with actuators, \(T_q\) is the time constant associated with drag, \(K_z\) is the static system gain, and \(I_3\) is the 3 by 3 square identity matrix. The states vector is given by \(\mathbf{x}_z=[x_1,x_2,x_3]^T=[z_I,\dot{z}_I,\ddot{z}_I]^T\) where \(z_I\) is the altitude in inertial frame. For what follows we simply drop all notation related to frame of reference as it is irrelevant for this contribution.

The sampled version of the system in Eq. \eqref{eq_state_space} is given by:
\begin{equation}\label{eq_ss_disc}
    \begin{split}
    \mathbf{x}_z(kh+h) &= \Phi(h) \mathbf{x}_z(kh) + \Gamma(h) u_z(kh)\\
    \mathbf{y}_z(kh)       &= C_z \mathbf{x}_z(kh) 
    \end{split}
\end{equation}
where \(k\in \mathbb{N}^+\), \(h\in \mathbb{R}^+\) is the sampling period, and \(\Phi\) and \(\Gamma\) are functions given by:
\begin{equation}
    \begin{split}
        \Phi(t)=e^{A_zt}\\
        \Gamma(t)=\int_0^t{e^{A_zs}B_zds}
    \end{split}
\end{equation}

Considering a time delay quantity given by \(\tau\) where it can be decomposed to:
\begin{equation}
    \tau=(d-1)h+\tau'
\end{equation}
where \(d\in \mathbb{N}^+\), and \(\tau'>0\) is the fractional delay. For convenience we are following the notation used by \cite{wittenmark1985sampling}. The sampled and delayed version of the system in Eq. \eqref{eq_ss_disc} is given by:
\begin{equation}\label{eq_ss_disc_delayed}
    \begin{split}
    \mathbf{x}_z(kh+h) &= \Phi(h) \mathbf{x}_z(kh) + \Gamma(h) u_z(kh-\tau_i)\\
    \mathbf{y}_z(kh)       &= C_z \mathbf{x}_z(kh-\tau_o) 
    \end{split}
\end{equation}
with the subscripts \(i\) and \(o\) associated with input and output delays respectively. The representation in Eq. \eqref{eq_ss_disc_delayed} does not represent all the states of the system since \(u_z(kh-\tau_i)\) and \(\mathbf{x}_z(kh-\tau_o)\) are not available in the states vector \(\mathbf{x}_z(kh)\). Thus we rewrite Eq. \eqref{eq_ss_disc_delayed} to be :
\begin{equation}\label{eq_ss_disc_delayed_ext}
    \begin{split}
    \mathbf{x}_{z,e}(kh+h) &= A_{z,e} \mathbf{x}_{z,e}(kh) + B_{z,e} u_{z,e}(kh)\\
    \mathbf{y}_z(kh)       &= C_{z,e} \mathbf{x}_{z,e}(kh) 
    \end{split}
\end{equation}
where the new extended state vector and state space matrices depend on \(\tau_i\) and \(\tau_o\). For example, for the case when \(d_i=d_o=2\) and \(\tau_i,\tau_o>0\) the extended state vector is given by:
\begin{gather*}
    \mathbf{x}_{z,e}(kh)=
    \begin{bmatrix}
       \mathbf{x}_z(kh) \\
       \mathbf{y}_{z,s}(kh)\\
       \mathbf{y}_{z,s}(kh-h)\\
       u(kh-2h)\\
       u(kh-h)
   \end{bmatrix}
\end{gather*}
and the system matrices for the case of \(\tau_o'>\tau_i'\) are given by:
\begin{gather*}
    A_{z,e}=
   \begin{bmatrix}
       \Phi(h) & 0 & 0 & \Phi(h-\tau_i')\Gamma(\tau_i') & \Gamma(h-\tau_i') \\
       \Phi(h-\tau_o') & 0 & 0 & \Phi(h-\tau_i')\Gamma(\tau_i') & \Phi(h-\tau_i'-\tau_o')\\
       0 & 1 & 0 & 0 & 0\\
       0 & 0 & 0 & 0 & 1\\
       0 & 0 & 0 & 0 & 0
   \end{bmatrix}\\
   B_{z,e}=
   \begin{bmatrix}
       0\\
       0\\
       0\\
       0\\
       1
   \end{bmatrix}\\
   C_{z,e}=   \begin{bmatrix}
       0 & 0 & I_3 & 0 & 0\\
       0 & 0 & 0 & 1 & 0\\
       0 & 0 & 0 & 0 & 1
   \end{bmatrix}\\
\end{gather*}
and for the case of \(\tau_o'<\tau_i'\), matrix \(A_{z,e}\) becomes:
\begin{gather*}
    A_{z,e}=
   \begin{bmatrix}
       \Phi(h) & 0 & 0 & \Phi(h-\tau_i')\Gamma(\tau_i') & \Gamma(h-\tau_i') \\
       \Phi(h-\tau_o') & 0 & 0 & 0 & \Phi(h-\tau_{io}')\Gamma(\tau_{io}')\\
       0 & 1 & 0 & 0 & 0\\
       0 & 0 & 0 & 0 & 1\\
       0 & 0 & 0 & 0 & 0
   \end{bmatrix}
\end{gather*}
where \(\tau_{io}'=\tau_i'-\tau_o'\). In general, the number of states of the system for the case when \(\tau_i,\tau_o>0\) is \(d_i+d_o+1\). Note that the case where \(\tau_i,\tau_o \equiv 0\) is impossible for cyber-physical systems.
\begin{remark}\label{remark_1}
A system with \(\tau_o>0\) is POMDP since \(\mathbf{y}(kh)\neq \mathbf{x}(kh)\)
\end{remark}

Based on Remark \ref{remark_1} a model-based state estimator that accounts for time delay, similar to the one developed in \cite{wahbah2022real}, is required to estimate \(\hat{\mathbf{x}}_z(kh)\). But for the case of \(\tau=\tau_i\) the POMDP system can be transformed to a MDP by simply selecting \(d_i\) previous inputs as states.

A difficulty associated with the system representation in \eqref{eq_ss_disc_delayed_ext} is that the selection of the number of states requires exact knowledge of both \(d_i\) and \(d_o\). Moreover, the extended states vector length can be substantially larger than the delay-free system if high sampling rate is used with systems with large delays. Such scenario is quite common in vision based robotics where inertial sensors sample at high rates while having large delays in the vision pipeline. In such case, we suggest the use of a larger sampling period \(\bar{h}\) that ensures \(\bar{h}>\tau\), which is possible practically and requires minimal number of additional states. In such case \(d_o=d_i=1\), and the extended states are given by:
\begin{gather}\label{eq_extended_states}
    \mathbf{x}_{z,e}(k\bar{h})=
    \begin{bmatrix}
       \mathbf{x}_z(k\bar{h}) \\
       \mathbf{y}_{z,s}(k\bar{h})\\
       u(k\bar{h}-\bar{h})
   \end{bmatrix}
\end{gather}
and the system matrices are given by (we are presenting the case of \(\tau_o'>\tau_i'\), refer to the aforementioned example):
\begin{gather*}
    A_{z,e}=
   \begin{bmatrix}
       \Phi(\bar{h}) & 0 & \Phi(\bar{h}-\tau_i')\Gamma(\tau_i')  \\
       \Phi(\bar{h}-\tau_o') & 0 & \Phi(\bar{h}-\tau_i')\Gamma(\tau_i') \\
       0 & 0 & 0
   \end{bmatrix}\\
   B_{z,e}=
   \begin{bmatrix}
       \Gamma(\bar{h}-\tau_i')\\
       \Phi(\bar{h}-\tau_i'-\tau_o')\\
       1
   \end{bmatrix}\\
   C_{z,e}=   \begin{bmatrix}
       0 & I_3 & 0\\
       0 & 0 & 1
   \end{bmatrix}\\
\end{gather*}

Since the state \(\mathbf{x}_z(k\bar{h})\) is not measurable, the system is characterized by the model dynamics and \(\mathbf{y}_z(k\bar{h}+\bar{h})\), \(\mathbf{y}_z(k\bar{h})\), and \(u_{z,e}(k\bar{h}-\bar{h})\) which is non causal, and hence the problem is non-Markovian. Therefore, the observed system states that would approximate the MDP to be used for RL training are given by:
\begin{gather}\label{eq_reduced_state}
    \hat{\mathbf{y}}_{z,e}(k\bar{h})=
    \begin{bmatrix}
       \mathbf{y}_{z,s}(k\bar{h})\\
       u(k\bar{h}-\bar{h})
   \end{bmatrix}
\end{gather}
Note that for the case when \(\tau_o'=0\) the POMDP converts to MDP since \(\mathbf{y}(k\bar{h})= \mathbf{x}(k\bar{h})\), and hence Eq. \eqref{eq_reduced_state} fully characterize the system state.
\begin{figure}[t]
\centering
\begin{subfigure}{0.35\textwidth}
        \includegraphics[width=\linewidth]{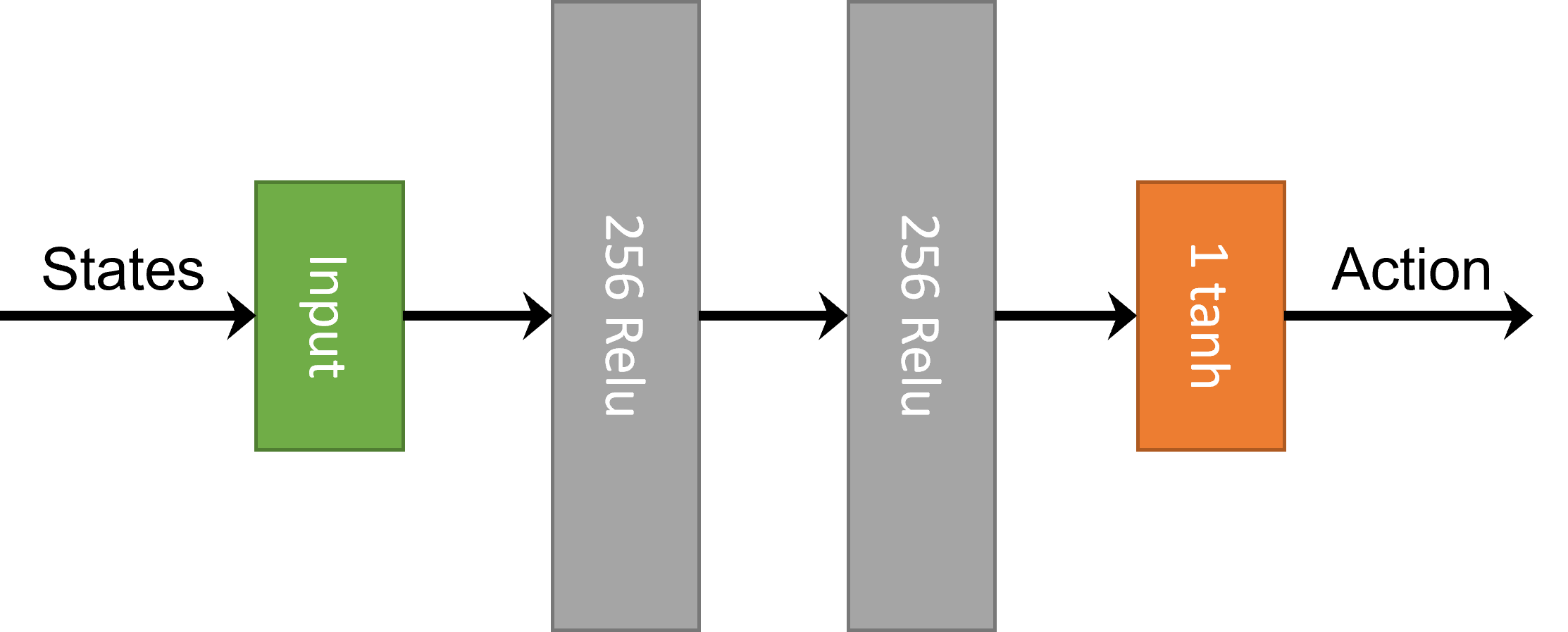}
        \caption{}
        \label{fig:PC}
    \end{subfigure}

\begin{subfigure}{0.45\textwidth}
        \includegraphics[width=\linewidth]{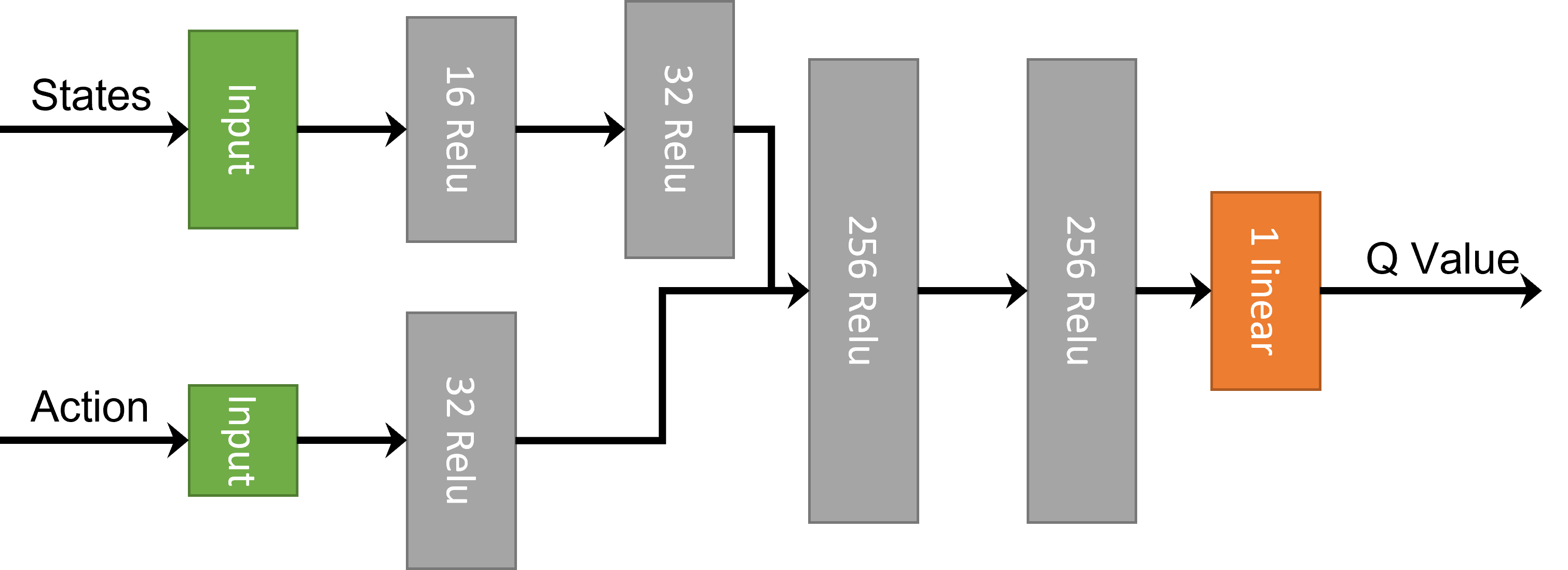}
        \caption{}
        \label{fig:PC}
    \end{subfigure}
\caption{The architectures of the actor network (a) and the critic network (b).}\label{fig_dnn_structure}
\end{figure}

\section{RL model}
A number of RL algorithms and techniques has emerged recently with successful deployment on various robotics applications \cite{tan2018sim,ibarz2021train,johannink2019residual,nguyen2019review}. In this work we use deep deterministic policy gradient (DDPG) RL algorithm \cite{lillicrap2015continuous} without loss of generality of the proposed contribution.

A few hyperparameters need to be selected for DDPG. The Polyak averaging parameter is chosen to be 0.005. The future reward discount factor used is 0.99. Fig. \ref{fig_dnn_structure} shows the details of the deep neural networks (DNN) used for both the actor and the critic. ADAM optimizer is used for loss minimization, with a learning rate of 0.001 and 0.002 for the actor and the critic networks respectively. We use experience replay for off-policy sampling with buffer size of 50,000 and a batch size of 1024. A noise model based on Ornstein-Uhlenbeck stochastic process is used to add random actions to the actor output during training for state space exploration. The RL setup is a modification of \cite{amifunny_2020}.

The actor model output is multiplied by \(u_{max}\) which represents the actuator saturation limit. This is done since the output layer of the actor model uses \(tanh\) as an activation function.

The reward function we used is given by:
\begin{equation}\label{eq_reward}
    r=-c||\mathbf{c}||_2
\end{equation}
where \(c\) is a scaling coefficient for numerical stability (we use \(c=100\)), and \(\mathbf{c}\) is the cost vector given by:
\begin{equation}
    \mathbf{c}=[z-z_{ref}\;\; \alpha \dot{z}]^T
\end{equation}
where \(\alpha\) is a weighing factor which we chose to be 0.1. In this work we define the RL task to be step following, so we simply choose \(z_{ref}=1\). We use step following for both training and testing of the agent.

\section{Results}
\subsection{Simulation Setup}
The model used for simulation represents the UAV altitude dynamics presented in Eq. \ref{eq_ss_disc} for the delay-free case and Eq. \ref{eq_ss_disc_delayed} for the delayed case. The simulation model was developed in MATLAB/Simulink software and has a fixed solver time step of 0.5 ms. The simulation program is compiled to produce an executable, which is interfaced with a Python program running a TensorFlow model of the RL algorithm.

The UAV altitude parameters we are using are \(T_p=0.049\) s, \(T_q=0.563\) s, \(K_z=0.84\), and a time delay of \(\tau=0.05\) s. These parameters correspond to real multirotor UAV altitude dynamics and they were found using the DNN-MRFT approach developed in \cite{Ayyad2020,ayyad2021tcst}. Such time delay figures were observed when using vision based measurements with heavy computational algorithms as we have demonstrated in our previous contributions \cite{hay2021unified,humais2022analysis}. DNN-MRFT approach is limited in the sense that it can only observe the total loop delay \(\tau\) and cannot distinguish \(\tau_o\) and \(\tau_i\). In the simulation we simply assume that all the delay is in \(\tau_o\) (unless stated otherwise), which is the most challenging case since the problem remains non-Markovian (refer to Remark \ref{remark_1}). The saturation limit that is used is \(u_{max}=6.57\)

The simulation time for every episode is more than 12 s to ensure stability during training. Each episode terminates after the simulation time is elapsed. We ran four simulation setups as follows:
\begin{itemize}
    \item \textbf{Case I}: observed states corresponding to the delay-free system in Eq. \eqref{eq_ss_disc} are used in RL training. Period of \(h=5\) ms is used for training.
    \item \textbf{Case II}: partially observed states corresponding to the delayed system in Eq. \eqref{eq_ss_disc_delayed} are used in RL training. These states are the \(\mathbf{y}_{z,s}(kh)\) states given in Eq. \eqref{eq_ss_disc_delayed_ext}. In this case \(\tau\equiv\tau_o'\) and \(\bar{h}=0.06\) s are used.
    \item \textbf{Case III}: extended state observation that corresponds to the vector in Eq. \eqref{eq_reduced_state} with \(\tau\equiv\tau_o'\) is used for RL training.
    \item \textbf{Case IV}: extended state observation that correspond to the vector in Eq. \eqref{eq_reduced_state} with \(\tau\equiv\tau_i'\) is used for RL training. This corresponds to the MDP case.
\end{itemize}

When training on models with delays, a sampling period larger than the time delay is used to yield the state vector in Eq. \eqref{eq_extended_states}. The sampling period used in Cases II, III, and IV corresponds to \(\bar{h}=0.06\) s.

\begin{figure}[t]
\centering
\begin{subfigure}{0.4\textwidth}
        \includegraphics[width=\linewidth]{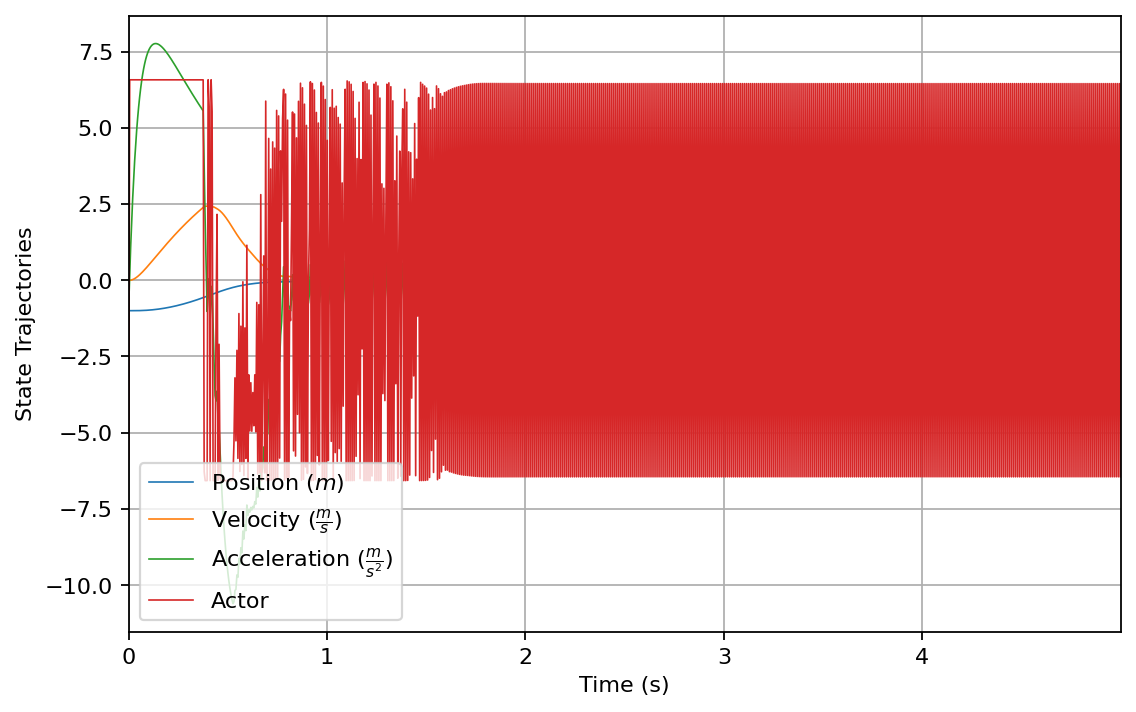}
        \caption{}
        \label{fig:case_1_a}
    \end{subfigure}

\begin{subfigure}{0.4\textwidth}
        \includegraphics[width=\linewidth]{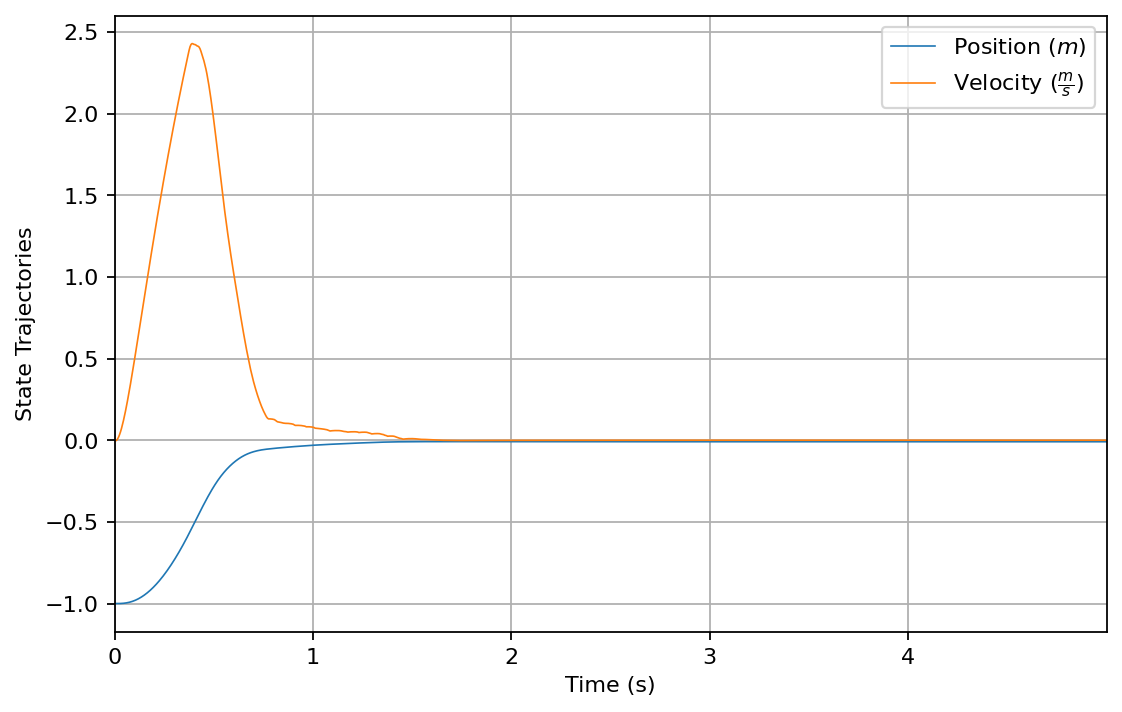}
        \caption{}
        \label{fig:case_1_b}
    \end{subfigure}

\begin{subfigure}{0.4\textwidth}
        \includegraphics[width=\linewidth]{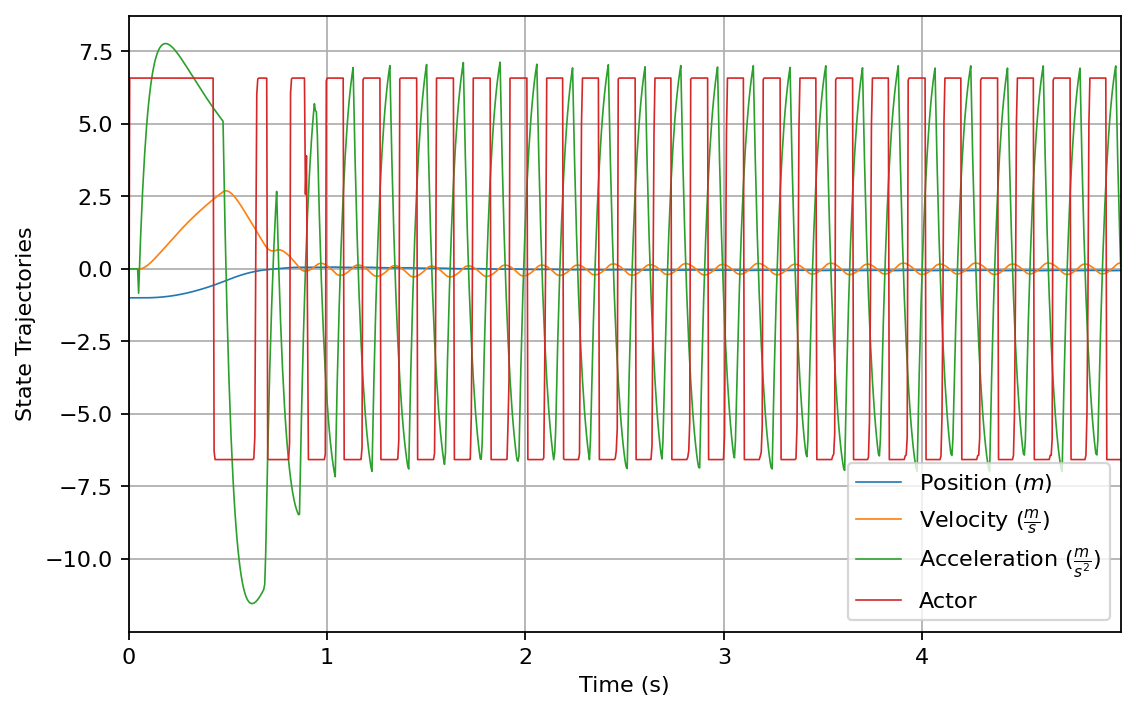}
        \caption{}
        \label{fig:case_1_c}
    \end{subfigure}
\caption{(a) Case I actor applied to a delay free system. (b) Same plot in (a) with acceleration measurement and actor output omitted for clarity. In (c) Case I actor is applied to a system with \(\tau=\tau_o=50\) ms.}\label{fig_case_I}
\end{figure}

\subsection{Effect of Delay Omission from the Model}
First we show that training an RL agent on a delay-free model, i.e. Case I, would result in \textit{incautious} actor that would yield oscillatory behaviour. Fig. \ref{fig_case_I} compares the step response of the RL agent trained in Case I when applied both to the delay-free simulation and the simulation with delay. Note that the oscillations in Fig. \ref{fig:case_1_a} are of high frequency, similar to chattering in sliding mode controllers \cite{boiko2005analysis}. The incautious actor in Case I is doing well in keeping position and velocity states at zero as seen in Fig. \ref{fig:case_1_b}, but this happens only in the ideal case of delay free systems. The excessive action generated by the actor is evident when the delay is introduced, which explains the oscillatory behaviour observed when transferring to reality as in \cite{tan2018sim}. Note that Case I actor is causing the actuator to alternate between positive and negative saturation limits as seen in Fig. \ref{fig:case_1_c}.

\begin{figure}
    \centering
    \includegraphics[width =0.8\linewidth]{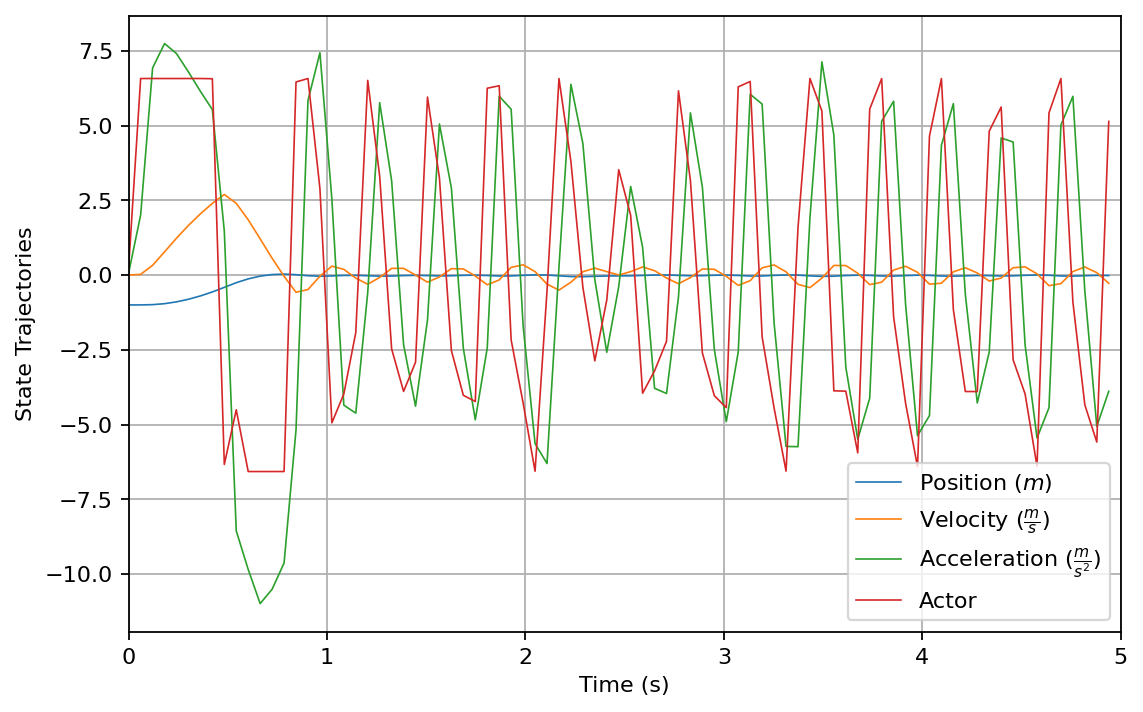}
    \caption{Case II actor when applied to the delayed model it was trained on. The actor results in large actions and causes noticeable velocity oscillations near zero.}
    \label{fig:case_2}
\end{figure}

On the other hand, training an RL agent on Case II yields a less oscillatory behaviour when applied to both delayed and delay-free systems. This behaviour is attributed to the fact that RL agents trained on delayed models are more conservative, and hence they do not fail when applied to delay-free systems. Fig. \ref{fig:case_2} illustrates Case II RL agent performance on the delayed system it was trained on. The RL actor in Case II produces oscillations with lesser amplitude and larger periods compared to Case I actor.

\begin{figure}[t]
\centering
\begin{subfigure}{0.4\textwidth}
        \includegraphics[width=\linewidth]{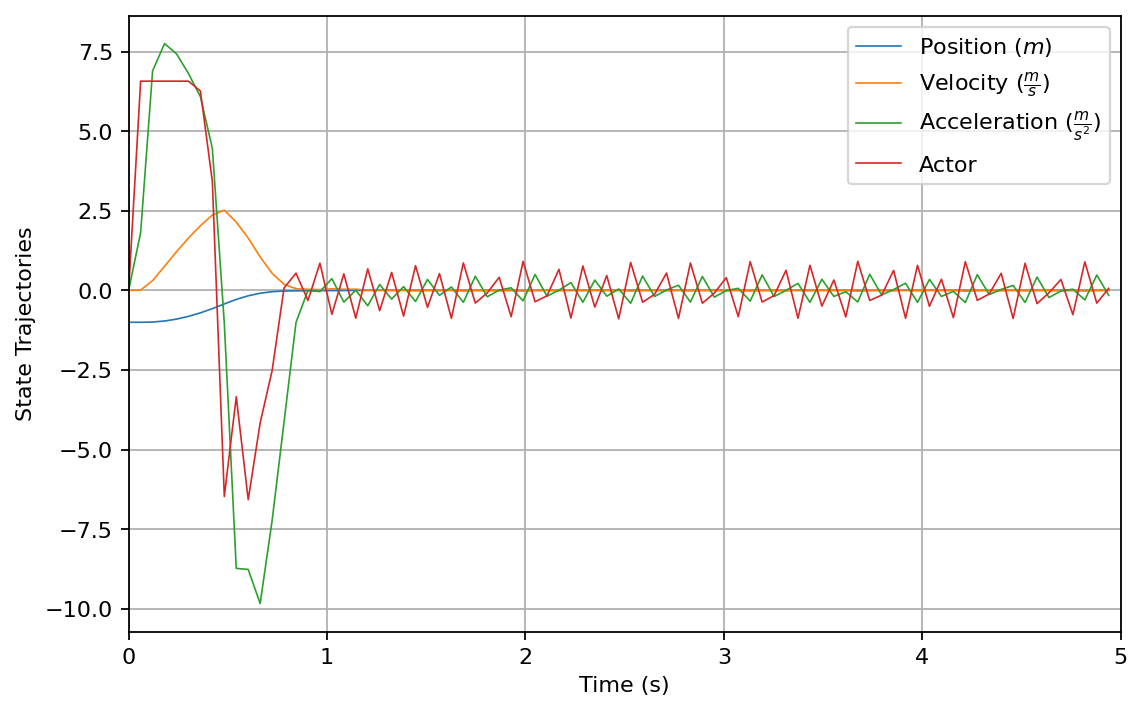}
        \caption{}
        \label{fig:case_3_a}
    \end{subfigure}

\begin{subfigure}{0.4\textwidth}
        \includegraphics[width=\linewidth]{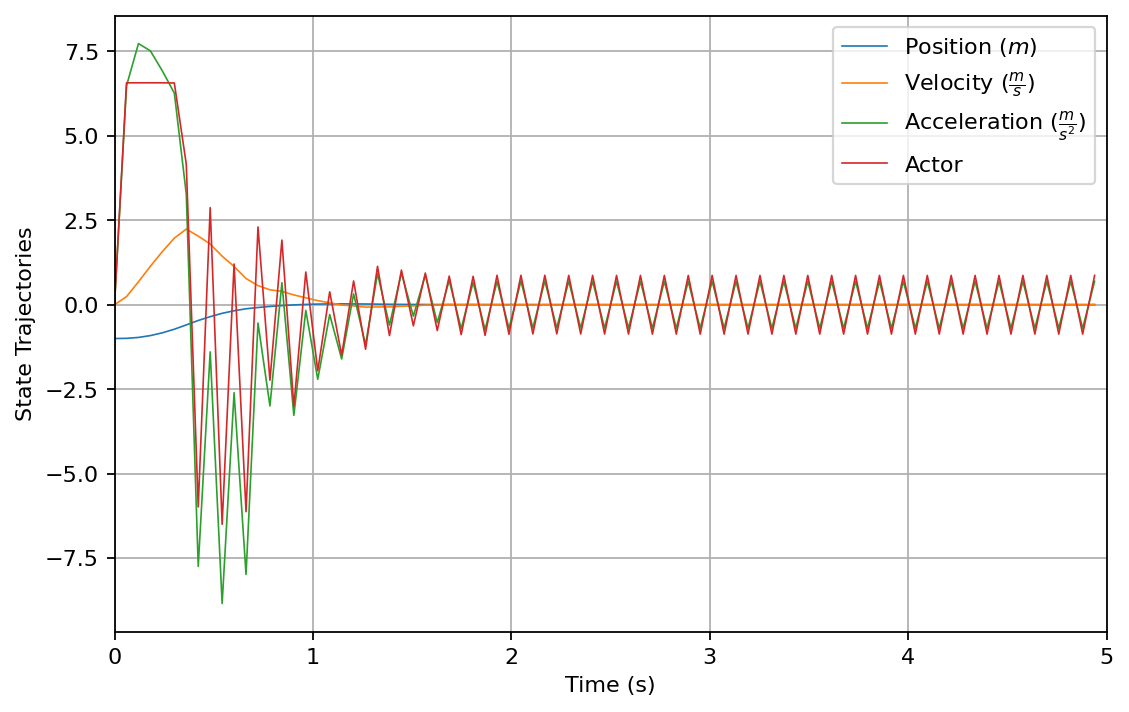}
        \caption{}
        \label{fig:case_3_b}
    \end{subfigure}

\caption{(a) Case III actor applied to a system with \(\tau_o'=50\) ms. (b) Case III actor applied to the delay free system. The cautious actor in Case III has more awareness about the state of the system compared to Case II, and thus results in much smaller oscillations.}\label{fig_case_3}
\end{figure}

\subsection{Training with Extended States Vector}
Case III and Case IV actors were trained using extended system state vector. Not much difference was noticed in the behaviour between Case III and Case IV actors despite that the former represents a POMDP and the later represents a MDP. This might be attributed to the fact that the physical model parameters used for training and testing are the same, and Case III actor and critic memorized, from repeated experience, the underlying dynamics. Thus we only report Case III results for convenience.

Case III and IV actors are \textit{cautious} compared to Case I actor. Fig. \ref{fig_case_3} shows Case III actor performance on both the delay and delay-free cases. The position and velocity states are almost zero as they are contributing explicitly to the RL reward presented in Eq. \eqref{eq_reward}. The cautious actor of Cases III and IV is not producing excessive actions as in Case I and II despite the fact that neither action nor acceleration are penalized. The cautiousness of Case III and IV actors is not a trade-off in performance since the rising time for all Cases I, II, III, and IV is almost the same. Thus the time delay involvement in model training, and the careful selection of the extended system state vector, both result in the correct behaviour required in practice.

It might be possible to use RL techniques like DR, DA, and reward hacking such that Case I and II would result in more cautious RL agents. The downsides of such approaches are plenty. First they introduce unintuitive hyper-parameters that do not correspond to explainable physical behaviour. As a result, the added hyper-parameters make training RL agents more expensive and difficult. Also these methods might result in over-conservative actors that do not perform well when deployed. The use of DR, DA, and reward hacking is therefore suggested to improve RL agent performance after correctly accounting for system delays and extending the states vector. 

\section{Conclusion}
This paper provided a theoretical investigation supported by simulation results about the effect of delays that are non-integer multiple of the sampling period on RL agents performance. It was shown that using a sampling period that is larger than the time delay and extending the state vector with just the previous action resulted in multiple improvements. First it reduced the RL training time significantly since the sampling period is enlarged and the number of states is almost the same as the delay free case. Second it produced RL agents that significantly outperformed agents trained on delay-free systems (Case I) or delay systems without considering the extended state vector (Case II). The reported results are preliminary, yet they provide insight about the research direction for successful use of RL for robotics. The results suggest that the more focus needs to be devoted to understanding the closed loop physical behaviour rather than exclusively focusing on unintuitive algorithmic upgrades. 





\bibliographystyle{IEEEtran}

\bibliography{bst/ref_items.bib}

\begin{IEEEbiography}[{\includegraphics[width=1in,height=1.25in,clip,keepaspectratio]{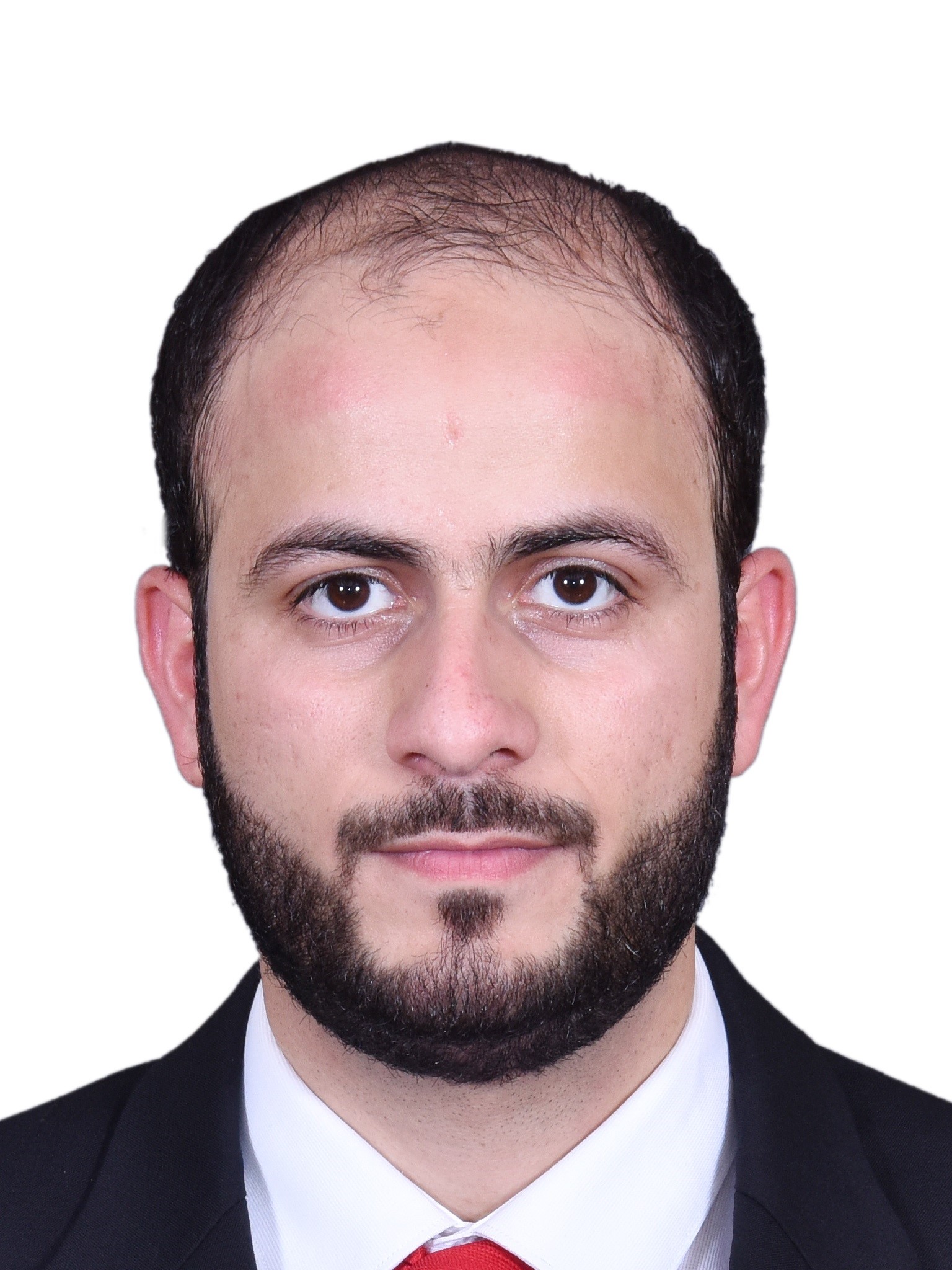}}]{Mohamad Chehadeh} received his MSc. in Electrical Engineering from Khalifa University, Abu Dhabi, UAE, in 2017. He is currently with Khalifa University Center for Autonomous Robotic Systems (KUCARS). His research interest is mainly focused on identification, perception, and control of complex dynamical systems utilizing the recent advancements in the field of AI.
\end{IEEEbiography}

\begin{IEEEbiography}[{\includegraphics[width=1in,height=1.25in,clip,keepaspectratio]{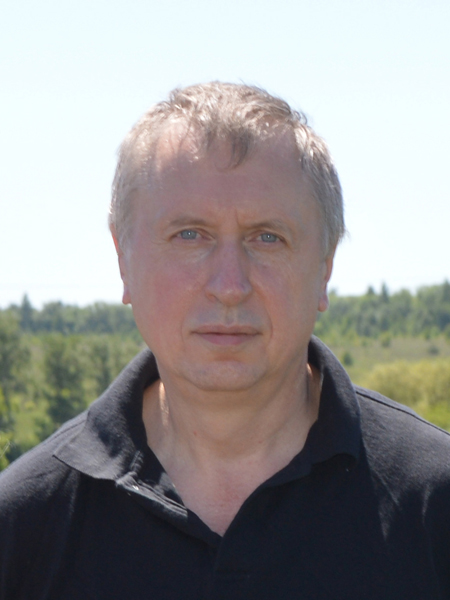}}]{Igor Boiko}
received his MSc, PhD and DSc degrees from Tula State University and Higher Attestation Commission, Russia. His research interests include frequency-domain methods of analysis and design of nonlinear systems, discontinuous and sliding mode control systems, PID control, process control theory and applications. Currently he is a Professor with Khalifa University, Abu Dhabi, UAE.
\end{IEEEbiography}

\begin{IEEEbiography}[{\includegraphics[width=1in,height=1.25in,clip,keepaspectratio]{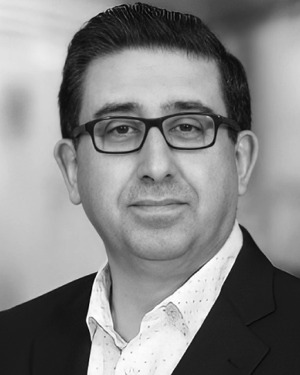}}]{Yahya Zweiri}
Yahya Zweiri (Member, IEEE) received the Ph.D. degree from the King’s College London in 2003. He is currently an Associate Professor with the Department of Aerospace Engineering and deputy director of Advanced Research and Innovation Center - Khalifa University, United Arab Emirates. He was involved in defense and security research projects in the last 20 years at the Defense Science and Technology Laboratory, King’s College London, and the King Abdullah II Design and Development Bureau, Jordan. He has published over 110 refereed journals and conference papers and filed ten patents in USA and U.K., in the unmanned systems field. His main expertise and research are in the area of robotic systems for extreme conditions with particular emphasis on applied Artificial Intelligence (AI) aspects and neuromorphic vision system.
\end{IEEEbiography}


\end{document}